\documentclass[aps,pra,twocolumn,10pt,showpacs]{revtex4}
\long\def\/*#1*/{}
\usepackage{float}
\usepackage{amsfonts}
\pdfoutput=1
\usepackage{graphicx}
\usepackage{amsmath}
\usepackage{amsthm}
\usepackage[colorlinks=true, citecolor=blue, urlcolor=blue ]{hyperref}
\usepackage{graphicx}
\usepackage{amsmath}
\usepackage{braket}
\usepackage{tikz}
\usepackage{appendix}
\usepackage{hyperref}
\usepackage{comment}
\usepackage{tikzpeople}
\usetikzlibrary{shapes.callouts}
\usepackage{bm}
\usepackage{IEEEtrantools}
\usepackage{xcolor}
\begin{document}

\title{Braess paradox in a quantum network} 
\author{Abhishek Banerjee, Pratapaditya Bej}
\email{banerjee.abhishek.official@gmail.com} \email{pratap6906@gmail.com}
\affiliation{Centre for Astroparticle Physics and Space Science, Bose Institute, EN 80, Sector V, Bidhannagar, Kolkata 700 091, India}

\begin{abstract}

Dietrich Braess while working on traffic modelling, noticed that traffic flow in a network can be worsened by adding extra edges to an existing network. This seemingly counter intuitive phenomenon is known as Braess paradox. 
 We consider a quantum network, where edges represent shared entangled states between spatially separated parties (nodes). The goal is to entangle two previously uncorrelated nodes using entanglement swappings. The amount of entanglement between the distant nodes is quantified by the average concurrence of the states established, as a result of the entanglement swappings. We then introduce an additional edge of maximally entangled Bell states in the network. We show that the introduction of the additional maximally entangled states to this network leads to lower concurrence between the two previously uncorrelated nodes. Thus we demonstrate the occurrence of a phenomenon in a quantum network that is analogous to the Braess paradox in traffic networks.

\end{abstract}

\maketitle

\section{Introduction}

	Average travel time for vehicles in a traffic network may increase upon adding extra roads to an existing network, as shown by Dietrich Braess in \cite{Ori_Braess,Eng_Trans}. Traffic flow in a network can be modelled as a strategic non-cooperative game, where the players(vehicles) are exposed to choices of different routes through the network, from the source node to the destination node. Rational choices on part of the vehicles maximize their individual payoff functions (minimize their travel times). Nash equilibrium \cite{Nash} for the network is the state where no vehicle can further reduce it's travel time by switching to another route, given that all the other vehicles stick to their choices of routes. The example shown below is taken from Braess original work.  
	
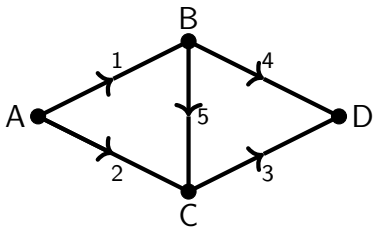
\begin{figure}
\centering
\begin{tikzpicture}
\draw [->] [black, ultra thick] (0,0) --(1,0.5);
\draw [black, ultra thick] (1,0.5) --(2,1);
\draw [->] [black, ultra thick] (2,-1) --(3, -0.5);
\draw [black,ultra thick] (3,-0.5) --(4,0);
\draw [->] [black,ultra thick](2,1) --(3,0.5);
\draw [black, ultra thick](3,0.5) --(4,0);
\draw [->] [black, ultra thick](0,0) --(1,-0.5); 
\draw [black, ultra thick](2,-1) --(0,0);
\draw [black, ultra thick](2,-1) --(2,0);
\draw [->][black, ultra thick] (2,1) --(2,0);
\draw [fill] (0,0) circle [radius=0.1];
\draw [fill] (2,1) circle [radius=0.1];
\draw [fill] (4,0) circle [radius=0.1];
\draw [fill] (2,-1) circle [radius=0.1];
\node [left, font = \large] at (-0.05,0) {$\mathsf{A}$};
\node [above, font = \large] at (2,1.05) {$\mathsf{B}$};
\node [right, font = \large] at (4.05,0) {$\mathsf{D}$};
\node [below, font = \large] at (2,-1.05) {$\mathsf{C}$};
\node [left] at (1.25,0.75) {$\mathsf{1}$};
\node [left] at (1.25,-0.75) {$\mathsf{2}$};
\node [left] at (3.25,-0.75) {$\mathsf{3}$};
\node [left] at (3.25,0.75) {$\mathsf{4}$};
\node [right] at (2,0) {$\mathsf{5}$};
\end{tikzpicture}
\caption{The traffic network consisting of four nodes labelled $A, \: B, \: C, \:  \text{and} \: D$ and five edges.}
\label{fig 1}
\end{figure}

	As shown in figure \ref{fig 1}, the network consists of $4$ nodes, $A$, $B$, $C$, and $D$. The edges are represented by $1$, $2$, $3$, $4$, and $5$. 


	 A total of $6$ vehicles are travelling from node $A$ to node $D$. The link travel times are linear functions of the number of vehicles using the links given  by: 
\begin{IEEEeqnarray}{rCl}
t_{1}=t_{3}=10x \nonumber \\ 
t_{4}=t_{2}=50+x \nonumber & & \text{  and} \\
t_{5}=10+x 
\end{IEEEeqnarray}
Initially, let's suppose the edge $5$ is absent, in that case the two paths ${ABD}$ and ${ACD}$ give equal travel time for each vehicle. Hence neither of the paths $ABD$ and $ACD$ is preferred to the other. We denote the number of vehicles on path `k' by $x_k$. The flow at Nash equilibrium, in this case is $x_{ABD} = x_{ACD} = 3.$ For this configuration, the average travel time is $83$ units.

	Now the network is modified by introducing the edge labelled $5$. In the presence of the edge $5$, the traffic distribution in the network, at Nash equilibrium  is $x_{ABD} = x_{ABCD} = x_{ACD} = 2$. The average travel time per vehicle for this configuration is $92$ units. This happens because, for some of the vehicles shifting to the edge $ABCD$ from their previously used path reduces their travel time, and hence presents itself as the rational alternative. Thus adding an extra edge to the network results in a deterioration in the performance of the network. Intuitively, adding extra resources to a network should increase the performance of the network, but as we saw in the example above, this is not always true. This seemingly paradoxical behaviour arises out of desire of minimizing individual travel times of each of the participants.
	
	Since then Braess paradox has been shown to occur in mechanical network of springs and strings \cite{Spring} where they have shown that in a network of strings and springs supporting a weight at equilibrium, cutting one of the strings involved, results in a new equilibrium where the weight rises.  In \cite{Meso_net}, it was shown that  a numerical simulation of quantum transport in a two-branch mesoscopic network reveals that adding a third branch can paradoxically induce transport inefficiency that manifests itself in a sizable conductance drop of the network. There are numerous other publications where Braess paradox has been shown to occur in basket ball games \cite{Basketball} and various other regimes. In \cite{QRG} it was shown that while sending classical information over a network, the effects of Braess paradox can be mitigated with access to quantum resources. In \cite{QDOT} Braess paradox was shown to occur in chaotic quantum dots.
	 
	It is evident that adding extra resource is not always beneficial to the overall performance of a network. This naturally leads to the question, whether such a paradoxical phenomenon can occur in the setting of a quantum network,  where shared entanglement between the nodes is a resource. We answer this question in the affirmative. 
	
	Advances in quantum information theory have made the realisation of the quantum internet a possibility. The Quantum internet is a network which interconnects remote quantum devices through quantum links along with classical ones. Quantum Networks is viewed as the natural supplement, if not the successor to the present day internet owing to the advantages quantum computing can provide in specific tasks \cite{Caleffi}.
	Shared quantum entanglement between spatially separated parties is one of the key resources in quantum information science and the backbone of quantum networks. It finds uses in various quantum informational tasks such as superdense coding \cite{SDC}, quantum key distribution \cite{BB84} etc. Distribution of entanglement between the spatially separated parties is a non-trivial problem in itself. Entanglement swapping \cite{swap1, swap2} is  one of the most widely used protocols for distributing entanglement between two remote parties, where quantum systems that have never interacted in the past, can nevertheless become entangled. Entanglement swapping has many applications in quantum networks. It can be used to enable the transmission of entanglement between long distances \cite{briegel,murali,su} and also used to distribute  quantum states  over arbitrary quantum networks \cite{clement, perseguers, perseguers2}. Entanglement swapping between Werner states and pure states has been studied in \cite{Kirby}.
	
	 We consider a quantum network of four nodes and four edges, as shown in \ref{fig 2}. Each of the nodes is in possession of a certain number of qubits. An edge represents shared entanglement between the nodes connected by the edge. Such networks have been studied extensively in \cite{cirac,sbroadfoot1,sbroadfoot2,acin,multper}, for a comprehensive review see \cite{perseguers2}. All the nodes are allowed to perform local operations on his/her qubits and can communicate classically with each other. Practically it is not always possible to generate pure entangled states, due to the presence of interaction with the environment, therefore some of the edges in a network may represent mixed entangled states. We initially configure the network such that the edges which represent pure states have $2N$ states such that those can be distilled to give $N$ maximally entangled states. This choice is motivated by the fact that the Nash equilibrium for the original configuration of the network where the entanglement in all the final states are maximised happens when each of the paths ACB and ADB admit $N$ swappings. The maximum happens when the pure states shared are maximally entangled. When the configuration of the network is changed, these must also be able to accommodate at most $2N$ swappings as the new edge disturbs the Nash equilibrium. The entanglement established between two previously uncorrelated after the entanglement swapping is dependent on the entanglement of the resource states used to perform the swapping. In case of a single entanglement swapping it is well known, that if the resource states used in the entanglement swapping are maximally entangled Bell states, it results in the maximum amount of concurrence in the state shared between the two furthest nodes. In this article we address the question: Is more shared entangled states between the nodes of a network, always beneficial to performance of the network? It turns out that this is not necessarily true. We show that introduction of additional entanglement in the form of maximally entangled Bell states in a quantum network, where the parties are non-cooperative, i:e each party performing the swappings is interested in increasing the entanglement in the resultant state and does not care about what happens to the swappings achieved by others, can lead to a lower average concurrence established in the final states established as a result of the entanglement swappings, between the two uncorrelated nodes. Thus we demonstrate the occurrence Braess paradox in the setting of a quantum network, revealing that increasing the amount of entanglement in a network is not always beneficial. 
The rest of the paper is organised as follows. We present our results in the section II and section III gives the conclusions and open problems.

\section{Results}	

\begin{figure}
\centering
\begin{tikzpicture}
\draw [->] [blue, ultra thick] (0,0) --(1,0.5);
\draw [blue, ultra thick] (1,0.5) --(2,1);
\draw [->] [blue, ultra thick] (2,-1) --(3, -0.5);
\draw [blue,ultra thick] (3,-0.5) --(4,0);
\draw [->] [red,ultra thick](2,1) --(3,0.5);
\draw [red, ultra thick](3,0.5) --(4,0);
\draw [->] [red, ultra thick](0,0) --(1,-0.5); 
\draw [red, ultra thick](2,-1) --(0,0);
\draw [fill] (0,0) circle [radius=0.1];
\draw [fill] (2,1) circle [radius=0.1];
\draw [fill] (4,0) circle [radius=0.1];
\draw [fill] (2,-1) circle [radius=0.1];
\node [left] at (1.25,0.75) {$\mathsf{(N)}$};
\node [left] at (1.25,-0.75) {$\mathsf{(N)}$};
\node[name = Dave, shape = dave, shirt=black, label=below:Dave]    at (2,-1.45) {};
\node[charlie, shirt=black, mirrored, hair=brown, label=below:Charlie]    at (2,1.8) {};
\node[name=alice, shape=alice, hair=red, shirt=magenta, label=below:Alice]    at (-0.5,0.0) {};
\node[bob,undershirt=black, mirrored, shirt=blue,label=below:Bob]    at (4.5,0) {};
\node[ellipse callout, align=left, draw,yshift= 1cm, xshift=-1.2cm,
callout absolute pointer={(alice.mouth)},
font=\tiny] {Hey Charlie,\\ we need only\\ N states,\\ let's distill.};
\node[ellipse callout, align=left, draw,yshift= -1.5cm, xshift=3.5cm,
callout absolute pointer={(Dave.mouth)},
font=\tiny] {Hey Bob,\\ we need only\\ N states,\\ let's distill.};
\end{tikzpicture}
\caption{The four node network, with blue(black) lines depicting pure entangled states $\ket{\Psi}$, red(gray) lines depicting Werner states}
\label{fig 2}
\end{figure}
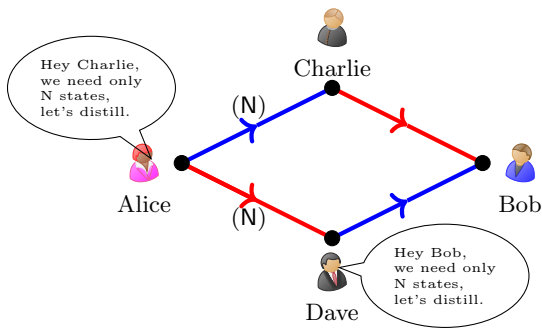

	We consider a network of four nodes and four edges, the edges represent shared entanglement between the nodes. Alice, Bob, Charlie and Dave are at the nodes as shown in figure \ref{fig 2}, henceforth referred to as, $A$, $B$, $C$, and $D$ respectively. Alice and Bob ($A$ and $B$) want to establish multiple entangled states between them which they can use later, such that the entanglement present in each of the states they share is maximized as permitted by the entanglement swappings permitted by the network. The number states they share was chosen to be $2N$ where $N$ is an integer. We chose this number to be even, so that at Nash equilibrium all the states established between $A$ and $B$ have the same amount of entanglement. This does not affect the results of this manuscript in any way.  We quantify entanglement using the measure concurrence \cite{Conc}. For entanglement swapping to be applicable there needs to be at least one intermediate node (say $C$). $ACB$ constitutes a path for entanglement swapping. For there to be another path of entanglement swapping we need at least another node ($D$). Now the two nodes $C$ and $D$ presents an option of introduction of an extra link in the network without affecting the uncorrelated nature of $A$ and $B$. So we have considered the simplest network configuration which allows the conditions of Braess paradox can occur.

The performance of the network is quantified by the average concurrence $C_{avg}$ of the $2N$ states established between $A$ and $B$. $C_{avg}$ is defined as 

\begin{equation}
C_{avg} = \frac{1}{2N}\sum_{i = 1} ^ {2N} C_i \ \text{where} \  i = 1, 2, ... , n
\end{equation}
	
Any pure quantum state of a composite system $AB$, $\Psi \in \mathcal{H}_A \otimes \mathcal{H}_B $ can be written as $\Psi = \sum_{i=1}^{n}{\lambda_i}\ket{i_A}\ket{i_B}$. where $\lambda_i$'s are non negative real numbers satisfying $\sum_{i=1}^{n}{\lambda_i^2} =1$, and $\ket{i_A}(\ket{i_B})$'s are orthonormal vectors in $\mathcal{H}_A(\mathcal{H}_B)$. These $\lambda_i$'s are known as the Schmidt coefficients of the state $\Psi $.
	
The parties connected by blue(black) lines have $2N$ two-qubit pure entangled states $\ket{\Psi}$ shared between them, where

\begin{IEEEeqnarray}{l}
\ket{\Psi_{AC}}=\ket{\Psi_{DB}}=\ket{\Psi} = \alpha\ket{00}+\sqrt{1-\alpha^2}\ket{11} \nonumber \\
\text{where} \: \alpha = \frac{1}{\sqrt[4]{2}}.
\end{IEEEeqnarray} 	

Where $\alpha$ and $\sqrt{1-\alpha^2}$ are the Schmidt coefficients of the states $\ket{\Psi}$.
	The concurrence of the states $\ket{\Psi}$ is given by:
\begin{equation}
C_{\Psi} = 2 \alpha  \sqrt{1-\alpha^2}
\end{equation}
$\alpha$ is chosen such that the $2N$ states can be deterministically transformed into $N$ Bell states using Nielesn's majorization criterion \cite{Nielsen}. The parties connected by red(gray) lines share $2N$ entangled Werner states \cite{Werner} given by:
\begin{multline}  
\rho_{AD} = \rho_{CB} = \rho_W = p\ket{\phi^+}\bra{\phi^+}+\frac{1-p}{4}{\mathcal{I}_4} \\
\text{where} \:  \frac{1}{3} \leq p \leq 1
\end{multline}
The concurrence of the Werner state $C_{W}$, is given by 
\begin{equation}
C_{W}=\frac{3p-1}{2}.
\end{equation} 
The states are chosen such that $C_{\Psi} > C_W$. Alice and Bob want to establish $2N$ shared entangled states $\mathbf{\rho}^{i}_{AB}\:\text{where} \: i = 1, 2, ..., 2N$ between them, for this they resort to Bob's and Charlie's help and ask them to perform entanglement swappings at their nodes, in a way such that the entanglement in $\rho_{AB}$, as quantified by concurrence is maximised for each ${i}$. To accomplish this task in the most efficient way possible, Alice and Charlie (Bob and Dave) perform deterministic entanglement distillation on their $2N$ states and prepare $N$ Bell states which are LOCC equivalent to 
\begin{equation}
\ket{\phi^+} = \frac{\ket{00} + \ket{11}}{\sqrt{2}},
\text{with concurrence} \: \: C_{\phi^+}=1
\end{equation}

No distillation is performed at the edges $AD$ and $CB$ since those are Werner states. Werner states can't be distilled deterministically to yield a predetermined number of states \cite{Puri}.

The concurrence of the final state $C_f$, after performing an entanglement swapping between a Werner state of concurrence $C_W$, and a pure state of concurrence $C_{\Psi}$ is given by \cite{Gour}
\begin{equation}
C_f = C_W C_{\Psi}
\end{equation}
Therefore average concurrence in $\rho^{i}_{AB}$ is 
\begin{equation}
C_{avg} = \frac{3p-1}{2}
\end{equation}

\begin{figure}
\centering
\begin{tikzpicture}
\draw [->] [blue, ultra thick] (0,0) --(1,0.5);
\draw [blue, ultra thick] (1,0.5) --(2,1);
\draw [->] [blue, ultra thick] (2,-1) --(3, -0.5);
\draw [blue,ultra thick] (3,-0.5) --(4,0);
\draw [->] [red,ultra thick](2,1) --(3,0.5);
\draw [red, ultra thick](3,0.5) --(4,0);
\draw [->] [red, ultra thick](0,0) --(1,-0.5); 
\draw [red, ultra thick](2,-1) --(0,0);
\draw [green, ultra thick](2,-1) --(2,0);
\draw [->][green, ultra thick] (2,1) --(2,0);
\draw [fill] (0,0) circle [radius=0.1];
\draw [fill] (2,1) circle [radius=0.1];
\draw [fill] (4,0) circle [radius=0.1];
\draw [fill] (2,-1) circle [radius=0.1];
\node [left] at (1.25,0.75) {$\mathsf{x+y}$};
\node [left] at (1.25,-0.75) {$\mathsf{z}$};
\node [dave, shirt=black, label=below:Dave]    at (2,-1.45) {};
\node [charlie, shirt=black, mirrored, label=above:Charlie]    at (2,1.45) {};
\node [alice, shirt=magenta, label=below:Alice]    at (-0.5,0.0) {};
\node [bob, mirrored, shirt=blue,label=below:Bob]    at (4.5,0) {};
\node [right] at (2,-0.1){$\mathsf{x}$};
\node [left] at (3.25,0.8) {$\mathsf{y}$};
\node [left] at (3.75,-0.8) {$\mathsf{x+z}$};
\end{tikzpicture}
\caption{The four node network, with blue(black) lines depicting pure non-maximally entangled states $\ket{\Psi}$, red(gray) lines depicting Werner states and green(light gray) line showing maximally entangled states, the number of swappings happening via the paths are depicted beside the edges by $x$, $y$, $z$, $x+y$,  and  $x+z$}
\label{fig 3}
\end{figure}
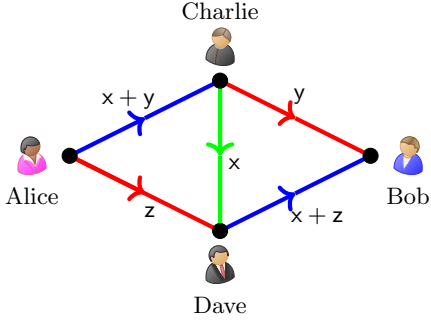

To see that this is the Nash equilibrium, assume Charlie performs $(N+1)$ entanglement swappings, so Dave performs $(N-1)$ entanglement swappings. Hence there has to be at least $(N+1)$ entangled states between Alice and Charlie. So the best they can do, in terms of concurrence is to distill from $2N$ states to $N+1$ states. By Nielesn's criterion \cite{Nielsen} the $(N+1)$ states they share after distillation have a concurrence $C'_{AC}$ given by
 
\begin{equation}
C'_{AC} = \left( \frac{1}{\sqrt{2}} \right) ^ \frac{N}{N+1} \cdot 
\sqrt{1- \left( \frac{1}{2} \right) ^ \frac{N}{N+1}  }  < 1
\end{equation}
The concurrence of the states established via these $N+1$ swappings is 
\begin{equation}
C_{{AB}_{(N+1)}} = C'_{AC}  C_W \leq C_W
\end{equation}
\textcolor{black}{The concurrence of the states shared between Bob and Dave is unchanged because they can get $N$ maximally entangled states. Dave uses $N-1$ states for swapping, while still having room to accommodate one more swapping via the path $DB$ with concurrence $1$. The one extra swapping via the path $ACB$ can increase the concurrence established by switching back to its original path $ADB$, as that will increase the concurrence.} Thus performing $N$ entanglement swappings each, which results in an average concurrence $C_{avg}$ in $\rho_{AB}$ given by:
\begin{equation}
C_{avg}=C_{\phi^+}  C_W = \frac{3p-1}{2}
\end{equation}
Now we modify the network by adding $2N$ maximally entangled states between Charlie and Dave, at the edge $CD$, shown by the green line in \ref{fig 3}, hoping that presence of more entangled states in the network, will lead to higher average concurrence in the final state $\rho_{AB}$ after the entanglement swappings. 

In this modified network Charlie (Dave) now has two options, he can perform a swapping between $\ket{\Psi_{AC}}$ \& $\rho_{CB}$ ($\rho_{AD}$ \& $\ket{\Psi_{DB}}$), or a swapping between $\ket{\Psi_{AC}}$ \& $\ket{\Phi^+_{CD}}$ ($\ket{\Phi^+_{CD}}$ \& $\ket{\Psi_{DB}}$). We call the the sequence of nodes involved in the entanglement swapping a path (eg. $ACDB$ is a path where Entanglement sawpping is performed between $\ket{\Psi_{AC}}, \ket{\Phi^+_{CD}}, \ket{\Psi_{DB}} $). If one swapping from the path $ADB$ now switches over to the path $ACDB$, the node $C$ has to accommodate $N+1$ swappings in total, so the edge $AC$ can now be deterministically distilled to $N+1$ non-maximally entangled states which have a concurrence 
\begin{equation}
C'_{AC} = \left( \frac{1}{\sqrt{2}} \right) ^ \frac{N}{N+1} 
\sqrt{1- \left( \frac{1}{2} \right) ^ \frac{N}{N+1} }
\end{equation}
The edge $DB$ still has $N$ maximally entangled pairs. The concurrence of the state resulting via the path $ACDB$ is 
\begin{equation}
C_{ACDB} = C'_{AC}  > C_\Psi  C_W.
\end{equation}
Clearly, choosing this path for the swapping provides an advantage. Therefore Charlie and Dave decide to perform more swappings along the path $ACDB$, for as long as this advantage over choosing $ACB$ or $ADB$ exists. The Nash equilibrium for this network is thus modified, and the average concurrence of the states $\rho^i_{AB}$, at Nash equilibrium is lower than the that of the original configuration. This is analogous to the Braess paradox in a traffic network where the introduction of an extra edge leads to an overall deterioration in the performance of the network. This behaviour is independent of $N$ as it always appears in the form of a ratio. The Braess paradox happens for for a wide range of values $p$.

To see this let's take an example with 
\begin{IEEEeqnarray}{l} 
N = 3 \nonumber \\ 
p = 0.9
\end{IEEEeqnarray}
In the original configuration of \ref{fig 1}, at Nash equilibrium, the average concurrence is $C_{avg}=0.8500$.

After adding the edge $CD$ suppose $x$ swappings are performed by the path $ACDB$, $y$ swappings are performed by the path $ACB$ and $z=6-(x+y)$ swappings are performed via the path $ADB$ as shown in \ref{fig 3}, Now the edges $AC$ and $DB$ are distilled according to the number of swappings they are participating in. If the altered Schmidt coefficients are denoted by $\alpha^{'2}_{AC}$ and $\alpha^{'2}_{DB}$ then:

\begin{equation*}
\alpha'_{AC} =  \left\{ \,
\begin{IEEEeqnarraybox}[][c]{l?s}
\IEEEstrut
\frac{1}{\sqrt{2}} & if $x+y \leq N$, \\
\left( \frac{1}{\sqrt{2}} \right) ^ \frac{N} {x + y} & if $x+y > N$.
\IEEEstrut
\end{IEEEeqnarraybox}
\right.
C'_{AC}=2\alpha'_{AC}\sqrt{1-\alpha^{'2}_{AC}}
\end{equation*}
\vspace{-5mm}
\begin{equation}
\alpha'_{DB} =  \left\{ \,
\begin{IEEEeqnarraybox}[][c]{l?s}
\IEEEstrut
\frac{1}{\sqrt{2}} & if $x+z \leq N$, \\
\left( \frac{1}{\sqrt{2}} \right) ^ \frac{N} {x + z} & if $x+z > N$.
\IEEEstrut
\end{IEEEeqnarraybox}
\right.\\
C'_{DB}=2\alpha'_{DB}\sqrt{1-\alpha^{'2}_{DB}}
\end{equation}

The average concurrence $C'_{avg}$ of the modified network for $x, y, \text{and} \  z$ where $x + y+ z = 2N$ is given by:

\begin{equation}
C'_{avg} = \frac{xC'_{AC}C'_{DB} + zC_{AD}C'_{DB} + yC'_{AC}C_{CB}}{2N} 
\end{equation}

The path $ACDB$ will be preferred to $ACB$ and $ADB$ for as long as 
\begin{equation}
C'_{AC}  C'_{DB} > C'_{AC} \cdot C_W
\end{equation}

Hence in this modified network at Nash equilibrium all the swappings happen via the path $ACDB$ and the average concurrence at Nash equilibrium is given by:

\begin{equation}
C'_{avg} = 0.8284
\end{equation}

Thus we see introducing additional resource to the network in the form of maximally entangled states, affects the performance of the network in  an adverse way as far as entanglement distribution is concerned. 

\begin{figure}
\begin{flushleft}
\includegraphics[scale=0.3]{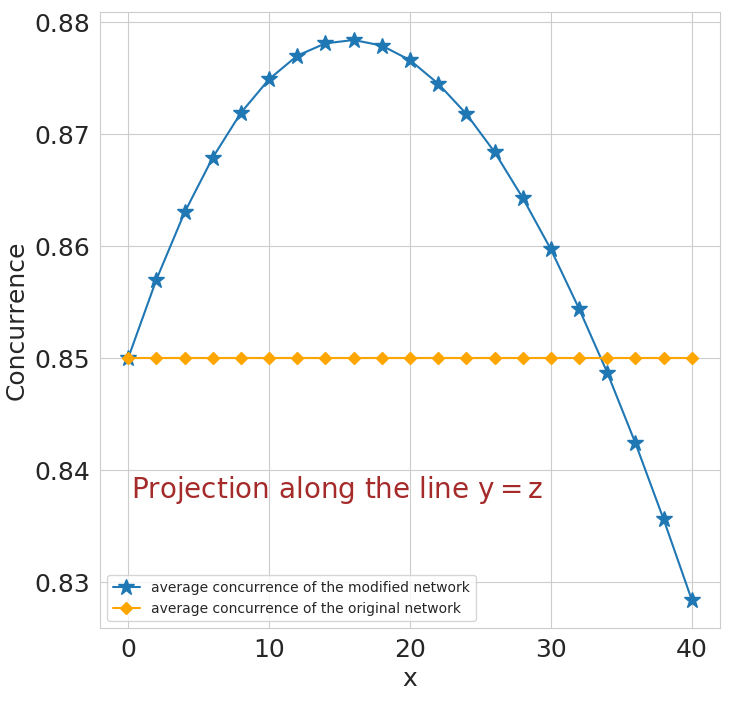}
\caption{Plot showing the variation of the average concurrence of the modified network in blue(dark gray) with $x$ projected along the line $y = z$, along with the average concurrence at Nash equilibrium, for the original network configuration shown in orange(light gray) for $N = 20$. Where $x$, $y$ and $z$ have the usual meanings from \ref{fig 3} and $x + y + z = 2N$. The line joining the points are for indicative purpose only and the variable $x$ is discrete.}

\label{fig 4}
\end{flushleft}
\end{figure}

In figure \ref{fig 4} we show the variation of average concurrence of the modified network shown in blue(dark gray) as a function of the number of swappings in the path $ACDB$ for $N = 20$ for $N = 20$. The value of $N$ is arbitrary and does not affect the nature of the plot as long as $N > 1$. Here we have introduced the constraint $y = z$ to first visualise a 2D plot. It shows that as $x$ increases, initially the average concurrence of the modified network increases to a maximum. Then as $x$ increases further and we move closer towards Nash equilibrium the average concurrence starts decreasing, and at Nash equilibrium, it falls below the average concurrence shown by the orange(light gray) plot of the network without the edge $CD$.

\begin{figure}[H]
\begin{flushleft}
\includegraphics[scale=0.36]{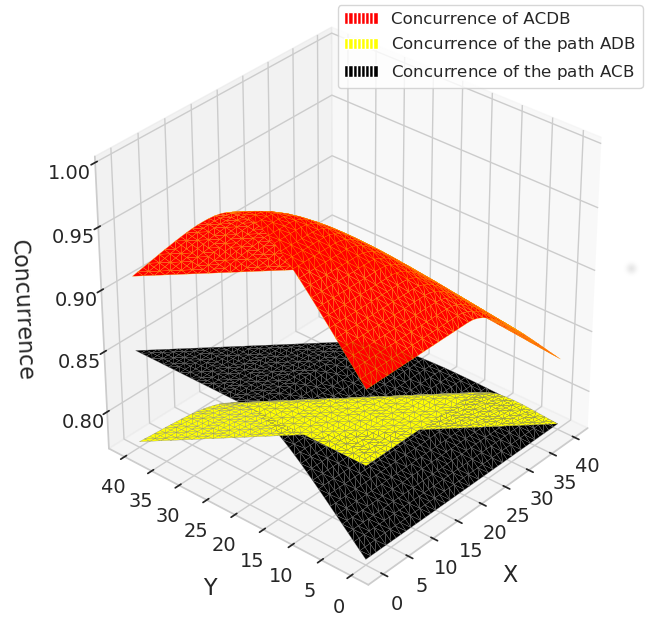}
\caption{Plot showing the variation of the concurrence of the path $ACDB$  shown in red(gray),  $ACB$ in black, and $ADB$ in yellow(light gray) with $x$ and $y$ for $N = 20$, where $x$ and $y$ have the usual meanings from \ref{fig 3}}

\label{fig 5}
\end{flushleft}
\end{figure}

\begin{figure}[H]
\begin{flushright}
\includegraphics[scale=0.38]{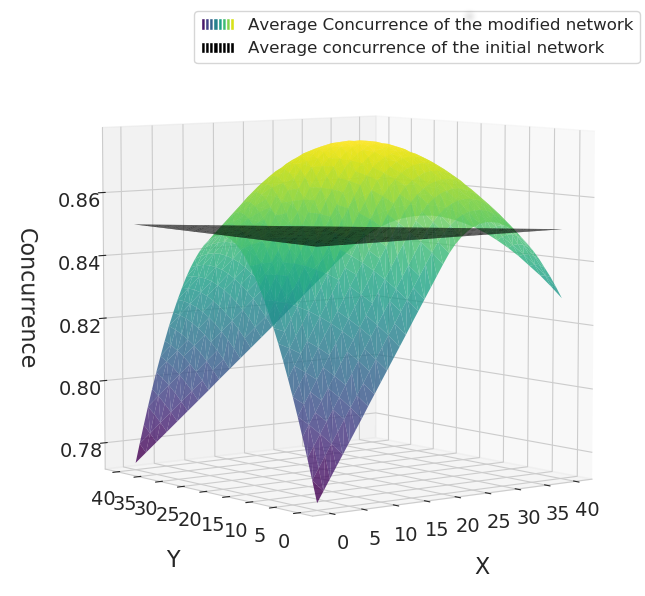}
\caption{Plot showing the variation of the average concurrence of the modified network shown in multi-colour(varying grayscale) with $x$ and $y$ along with the average concurrence at Nash equilibrium for the original network configuration shown in black for $N = 20$. Where $x$ and $y$ have the usual meanings from \ref{fig 3}}

\label{fig 6}
\end{flushright}
\end{figure}

\begin{figure}[H]
\begin{flushright}
\includegraphics[scale=0.34]{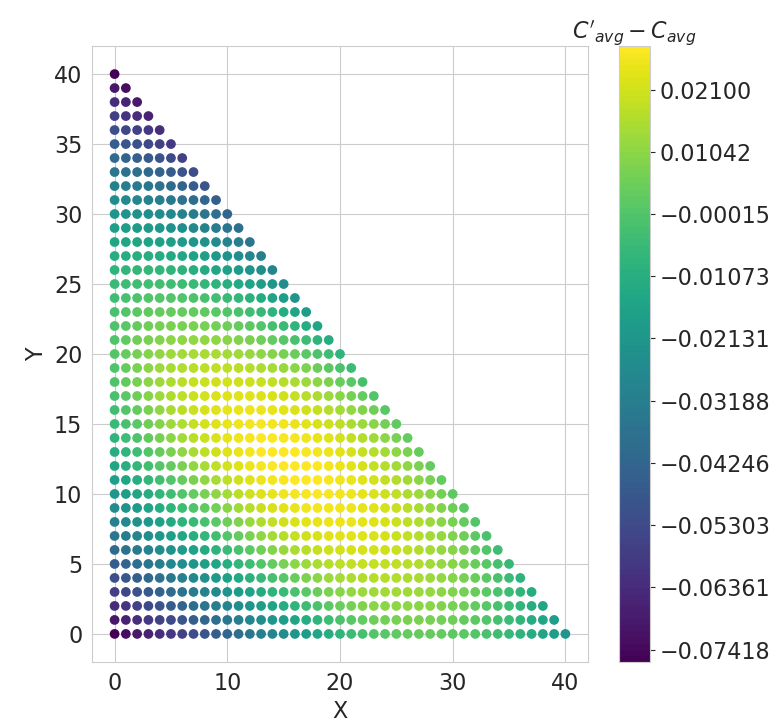}
\caption{Plot showing the variation of difference between the average concurrence of the modified network and the average concurrence of the original network configuration for $N = 20$. Where $x$ and $y$ have the usual meanings from \ref{fig 3}}

\label{fig 7}
\end{flushright}
\end{figure}

\begin{figure}
\begin{flushleft}
\includegraphics[scale=0.3]{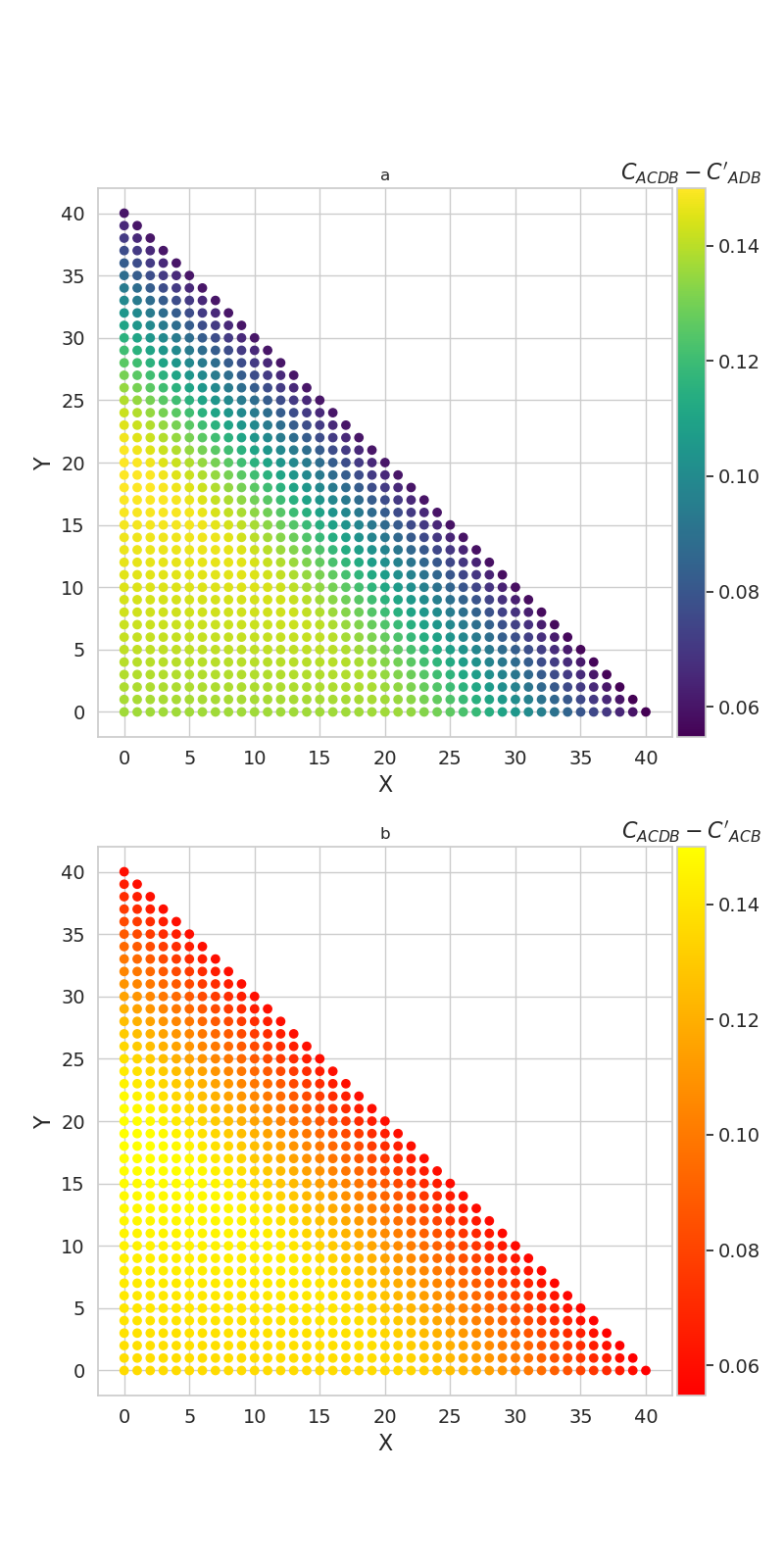}
\caption{Plot 8(a) shows the variation of difference between the concurrence of path $ACDB$ and $ADB$. Plot 8(b) shows the variation of difference between the concurrence of path ACDB and ACB for $N = 20$. where $x$ and $y$ have the usual meanings from \ref{fig 3}}

\label{fig 8}
\end{flushleft}
\end{figure}

Figure \ref{fig 5} shows that the concurrence resulting from swappings via the newly added path $ACDB$ shown in red(gray) always stays higher than the concurrence resulting via the other two paths viz. $ACB$ shown in \textcolor{blue}{black} and $ADB$ shown in yellow(light gray) for $N = 20$. There is always an incentive for every swapping to switch over to the path $ACDB$ for all values of $x$ and  $y$. 

In figure \ref{fig 6} we plot the average concurrence of the modified network shown in multi colour(varying grayscale) as a function $x$ and $y$ along with average concurrence of the initial configuration shown in black at Nash equilibrium for $N = 20$. 

Figure \ref{fig 7} shows the difference between the average concurrence of the modified network and the average concurrence of the initial network as a contour plot.

In Figure \ref{fig 8} we have plotted the difference in the concurrence of the path $ACDB$ and $ADB$ in the upper subplot and difference in the concurrence of $ACDB$ and $ACB$ in the lower plot.  We can see that as the difference between the concurrence of the path $ACDB$ and $ADB$ remains positive for all values of $x$ and $y$, the advantage to switch over to $ACDB$ from $ADB$ always exists. The same happens for the paths $ACDB$ and $ACB$ as well. It can be seen that as more swappings happen via the newly added path $ACDB$ the average concurrence of the network increases to a maximum and at that point, and then starts decreasing as more swappings continue to switch over to this path. At Nash equilibrium all the swappings shift to $ACDB$ the average concurrence falls below that of the original configuration of the network. The value of $N$ was chosen arbitrarily as it doesn't have any effect on the results.

 In essence, introduction of the maximally entangled states between the nodes $C$ and $D$, leads to poorer performance of the network, in spite of there being more entanglement available in the network.

\section{Discussion and conclusion}

We considered a four node quantum network $ABCD$ where $AC$, $BD$  share $2N$, similar two-qubit, non-maximally entangled, pure states and $AD$, $CB$ share $2N$, two-qubit mixed entangled states. $A$ and $B$ want to establish $2N$ entangled states between them using entanglement swapping where the entanglement in each of the states established is maximised.  We quantify the performance of this network using the average concurrence of the $2N$ states, established between nodes $A$ and $B$ as the figure of merit. We then introduce additional entanglement in the network, between the nodes $C$ and $D$ in the form of maximally entangled states, hoping this might lead to a better performance of the network, because in general, it is believed that increasing the resources might lead to an increase in the performance.

We have considered the scenario where each of the swappings try to maximise the entanglement established in the final state established after the swapping. As it might happen that Alice and Bob want to share entangled states between them such that all the states are equally useful in terms of the entanglement present in them.

When not at Nash equilibrium, the swappings via the paths $ACB$ and $ADB$ are not performing optimally, those swappings can benefit the most by switching over to the newly added path ACDB. As we have shown in figure 5, Initially as more swappings happen via the path $ACDB$ the average concurrence of the network increases. The state of the network where the average concurrence attains a maximum value, either the swappings via $ACB$ or $ADB$ can still increase the entanglement established via swapping by switching over to the path $ACDB$. This advantage on switching over to the newly added path $ACDB$ exists for as long as the network doesn't reach Nash equilibrium. Parties  $C$  and  $D$  are  acting  as  non-cooperative  agents  in  this  scheme  as  $C$  is concerned only with increasing the concurrence achieved by his swapping, and does not care about the effects it has on the swappings of $D$. Here the incentive is to maximise the entanglement established via the swapping. In such a scenario where all the swappings try to increase the entanglement in the resulting state, the quantum network will always try to gravitate towards the Nash equilibrium.

Our results show that in the current setting, if one tries to maximise the entanglement established in the states resulting from every swapping, the amount of entanglement in the final states is not maximised. The network performs best when some of the entanglement swappings settle for final states, in which the entanglement is not maximum as allowed by the available swapping options and can be increased by switching over to the newly added path. In the current setting additional entanglement introduced between arbitrary nodes could worsen the entanglement distribution between the intended nodes. In quantum networks, communication between two nodes might require the exchange of quantum information among the nodes. Quantum teleportation is one of the most widely used protocols to transfer quantum information between two spatially separated entangled nodes, without the need to physically transfer the qubit. The concurrence of an entangled state is in direct correspondence with the teleportation fidelity that can be achieved using the state as the resource \cite{horodecki, verstrate}. So in a way maximising the average concurrence renders the states established between Alice and Bob most useful for teleportation. It shows that even though maximally entangled states are useful resources for entanglement swapping at the individual level, in a network of multiple nodes and edges, extra entanglement might not be always profitable for the overall performance of the network.

\textcolor{black}{The importance of this result stems from the fact that, shared entanglement is one of the most important resources in quantum information processing, it facilitates many quantum informational tasks such as teleportation\cite{Tele_ori}, QKD \cite{BB84} etc. Entanglement swapping is one of the most widely used protocols to distribute entanglement between distant nodes. In spite of maximally entangled states being the ideal resource for entanglement swapping, extra maximally entangled states in a network can lead to lower entanglement between the intended nodes. In a quantum network if two distant parties want to establish entanglement between them via swapping, they can't rely on the intermediate nodes to choose the best path for maximizing the entanglement between them.} Braess paradox plays an important role in the design of classical networks, we have shown that the paradoxical behaviour can also arise in case of quantum networks, therefore the Braess paradox should be taken into consideration in the design of quantum networks as well

Although our findings are somewhat restricted by the structure of the network, just as in the case of the classical Braess paradox, it doesn't rule out the possibility of occurrence of the paradox in more complex network configurations. We have left this as an open question.

  The implications of this in the setting of complex quantum networks and entanglement percolation in higher dimensional networks could be interesting questions to investigate.
  
\textit{Acknowledgment}: The authors are grateful to Prof. Somshubhro Bandyopadhyay for helpful discussions.

\end{document}